\documentclass[conference]{IEEEtran}
\IEEEoverridecommandlockouts
\usepackage{amsmath,amssymb,amsfonts}
\usepackage{algorithmic}
\usepackage{graphicx}
\usepackage{textcomp}
\usepackage{xcolor}
\newcommand{\ml}{\textsc{ml}}
\newcommand{\mybl}{\color{black}}  
\begin{document}

\title{Blockchain Enabled Trustless API Marketplace}

\author{
\IEEEauthorblockN{Vijay Arya}
\IEEEauthorblockA{IBM Research \\ vijay.arya@in.ibm.com}
\and
\IEEEauthorblockN{Sayandeep Sen}
\IEEEauthorblockA{IBM Research \\ sayandes@in.ibm.com}
\and
\IEEEauthorblockN{Palani Kodeswaran}
\IEEEauthorblockA{IBM Research \\ Palani.kodeswaran@in.ibm.com}
}

\maketitle

\begin{abstract}
There has been an unprecedented surge in the number of service providers offering a wide range of machine learning prediction APIs for tasks such as image classification, language translation, etc. thereby monetizing the underlying data and trained models. Typically, a data owner (API provider) develops a model, often over proprietary data, and leverages the infrastructure services of a cloud vendor for hosting and serving API requests. Clearly, this model assumes complete trust between the API Provider and cloud vendor. On the other hand, a malicious/buggy cloud vendor may copy the APIs and offer an identical service, under-report model usage metrics, or unfairly discriminate between different API providers by offering them a nominal share of the revenue. In this work, we present the design of a blockchain based decentralized trustless API marketplace that enables all the stakeholders in the API ecosystem to audit the behavior of the parties without having to trust a single centralized entity. In particular, our system divides an AI model into multiple pieces and deploys them among multiple cloud vendors who then collaboratively execute the APIs. Our design ensures that cloud vendors cannot collude with each other to steal the combined model, while individual cloud vendors and clients cannot repudiate their input or model executions.
\end{abstract} 

\begin{IEEEkeywords}
Innovative use cases which leverage blockchain, AI, Model inference, Software as a service API
\end{IEEEkeywords} 

\section{Introduction}
\label{sec:introduction}


Last few years have witnessed a significant growth in cloud based API marketplaces that offer a diverse collection of software APIs on a pay-per-query basis~\cite{cloudai, mashape, rapidapi, cloudsight, azure, aws, clarifai}. Commonly, these APIs expose valuable software or business intelligence functions that application developers can integrate with data and quickly compose and build rich software applications.  More recently, with the growth and interest in building AI applications, most marketplaces offer a large collection of pretrained prediction APIs. These APIs span multiple domains such as social media, finance, healthcare, education, and advertising and can help perform tasks such as image tagging, face recognition, document classification, speech recognition, and sentiment analysis. For instance, a text analytics API may accept unstructured text data as input and detect its language or offer language translation service. A banking API may output a credit risk score given the historical transaction data of a customer. A healthcare API may assess the risk of cancer in a patient given the radiology data. The Algorithmia marketplace~\cite{algorithmia} hosts a library of about $4,500$ such algorithm APIs while the Mashape marketplace (part of RapidAPI) ~\cite{mashape, rapidapi} offers more than $10,000$ APIs. 


The API marketplace ecosystem generally consists of three stakeholders: a centralized \emph{cloud vendor}, \emph{API providers}, and \emph{API consumers} (i.e. application developers). 
The centralized cloud vendor (e.g. Mashape, Algorithmia, etc.) hosts the APIs developed by individual or small scale API providers and manages the infrastructure and compute resources needed to execute API query requests from consumers while guaranteeing security, availability, and low latency. The API providers develop predictive or prescriptive models that are commercially valuable and wrap these in the form of API containers~\cite{kub} that are hosted on the cloud platform. These models typically encompass proprietary knowledge regarding certain processes or features, which when run on specific data can produce output of economic value to end-users. Being part of a marketplace helps individual and small scale API providers to increase the accessibility and exposure of their APIs to a larger customer base and avoid challenges pertaining to building and hosting API portals. Lastly, the API consumers or application developers access the hosted APIs typically on a pay-per-query basis based on a subscription and pricing plan. The API consumers benefit from the marketplace ecosystem by having an increased choice and diversity of discoverable APIs and pricing plans to meet their software development requirements, thus lowering the overall cost of application development. The revenue generated from API calls is shared between the cloud vendor and the API providers. 
By connecting the API providers and consumers, the marketplace enables all three stakeholders to derive economic value from the joint ecosystem. 

{\mybl
\noindent\emph{Example.~} Consider a machine learning developer who trains a language translation model and wishes to monetize it. She therefore wraps her model within an API that accepts unstructured text and target language as inputs and returns the translated text. A developer generally lacks the infrastructure to host and serve APIs at scale, and therefore chooses a cloud vendor who hosts a centralized API marketplace (for instance  Algorithmia~\cite{algorithmia}), and deploys the API on the cloud. In addition to hosting APIs, some cloud vendors also provide automated services to convert ML models into APIs, which minimizes an AI developer's effort in monetizing her models. A web developer, for example responsible for developing the Wimbeldon website, integrates the API call via a translate button on the webpage, enabling users from different countries to view it. In this example, the web developer who is one of the API consumers, pays the cloud vendor based on the number of API calls per month. The cloud vendor in turn shares the API revenue with the ML developer, who is the API provider. }

\begin{figure}
    \centering
\includegraphics[width=\linewidth]{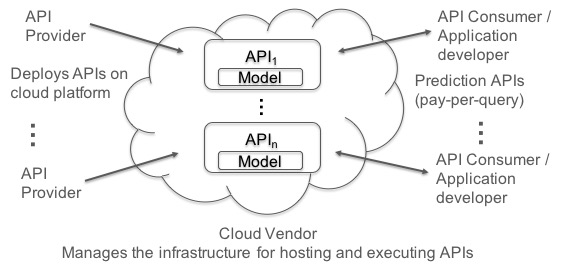}
    \caption{API Marketplace Ecosystem with three stakeholders: A centralized cloud vendor, API providers, and API consumers.}
    \label{fig:market}
\end{figure}

Despite providing numerous advantages, these centralized API marketplaces face challenges related to fairness and trust between all three stakeholders. More specifically, 
\begin{itemize}
\item \emph{A malicious cloud vendor} may copy a deployed API including the associated confidential model supplied by a provider and then offer a similar service. Moreover the cloud vendor can invoke the API multiple times without actually reporting it to the provider or report inaccurate API usage metrics to increase its own share of the API revenue. 

\item \emph{A malicious API provider} may repudiate the query results of a certain version of the deployed API or provide a wrong API to begin with and repudiate that the output has been generated by his API. The API provider may also blame the cloud vendor for not executing an accurate version of his API. 

\item \emph{A malicious API consumer} may demand refund by claiming that a wrong output was provided by the API service by possibly forging its output. The API consumer may first give wrong input data to the API and then claim that the output was wrong by claiming to have provided some other data as input.
\end{itemize} 

In this work, we present the design of a blockchain enabled decentralized API marketplace that ensures fair and trusted usage of hosted APIs and averts malicious behavior of all three stakeholders -- cloud vendor, API provider, and API consumer. The key novelty of the proposed solution lies in the fact that instead of having one central cloud vendor store and execute the confidential models encapsulated within an API, our system splits the storage and execution of models across multiple collaborating cloud vendors such that no single vendor has full knowledge of the models invoked during an API request. The system is designed to incentivize all three stakeholders to record transactions related to their actions on blockchain. All transactions related to the storage of models components across cloud vendors, invocation of API request by API consumers, execution of specific model components held by individual cloud vendors, and receipt of the final output by API consumers are stored on the Blockchain. The system hosts smart contracts to automatically enforce correct execution of functionality and provide evidence for dispute resolution by a trusted arbitrator. 

\begin{figure*}
    \centering
\includegraphics[width=0.48\linewidth]{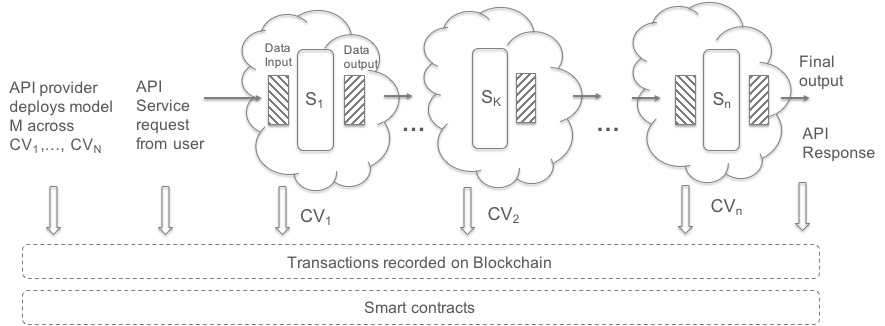}
\includegraphics[width=0.48\linewidth]{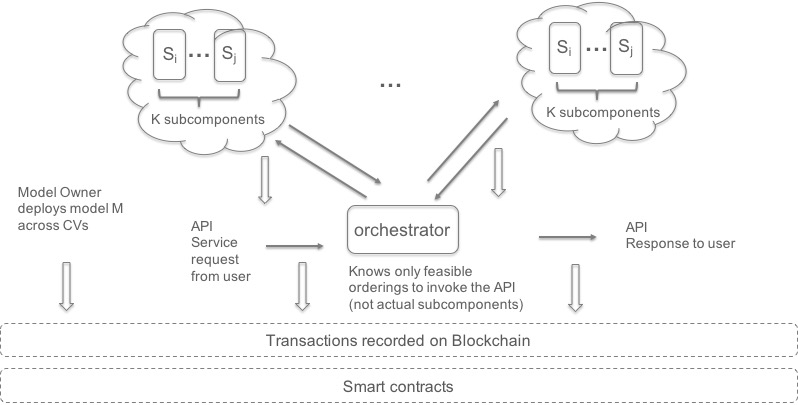}
    \caption{Blockchain enabled decentralized API Marketplace: API model invocations are split across multiple cloud vendors so that no single vendor has full knowledge of the underlying API model. All transactions are recorded by participants on the distributed ledger. (Left) Regular approach without orchestrator, (Right) Alternate Approach with orchestrator. }
    \label{fig:system}
\end{figure*}

\subsection{Related Work}
A number of centralized and decentralized marketplaces have emerged that provide different services related to training and deploying machine learning (\ml) models. These include marketplaces based on rewards, that enable training of \ml\ models on data held by data owners, purchase of data assets, and services to offer trained models as APIs. 

Kaggle is an example of centralized machine learning marketplace that connects data owners with machine learning developers and rewards developers that submit a model with the highest accuracy. The Danku protocol~\cite{danku} by Algorithmia offers a similar decentralized service, which uses blockchain. Using a smart contract, the data owners publish their data and an evaluation function on the blockchain. The developers utilize this data and submit trained models that are automatically tested using the evaluation function by miners on the blockchain network to announce a winner. 

The Skychain~\cite{skychain} project aims to connect owners of medical data with AI developers and medical professionals on blockchain. The project proposes the concept of proof of deep learning training, which enables crypto miners to  contribute GPU compute resources for training neural network models. Neureal is a similar effort~\cite{neureal} that allows idle computing power to be commoditized and used for analyzing big data, while rewarding miners for the accuracy of their predictions. Similarly, Deepchain~\cite{deepchain} attempts to achieve privacy-preserving model training, wherein mistrustful parties are given incentives to participate in federated learning by sharing gradients and correctly updating parameters. The Ocean protocol~\cite{ocean} is a blockchain based decentralized data exchange protocol, which connects data owners with data consumers (i.e. AI model developers). Owners can sell their data multiple times in a secure and transparent manner and a crowd based reputation system helps price the datasets in the marketplace. 

This work focuses on the deployment of trained models as APIs. While a number of centralized API marketplaces exist~\cite{cloudai, mashape, rapidapi, cloudsight, azure, aws, clarifai}, there is no prior work on blockchain based decentralized API marketplaces that connect API providers, consumers, and cloud vendors. Moreover, none of the existing marketplaces are based on the concept of collaborative execution of trained models by multiple cloud vendors. Additionally, unlike some of the existing work~\cite{skychain}, our work does not propose modifications to the underlying blockchain protocols and can function over different types of networks (e.g. Ethereum~\cite{ethereum, ethereum1} and Hyperledger fabric~\cite{hyper}). {\mybl Also, our work does not store API models on blockchain, which may lead to loss of model privacy.} Lastly, our work is more broadly applicable to a large variety of APIs that encapsulate different types of confidential models. These models may simply execute software code, a mathematical function, or invoke a trained \ml\ models for inference (e.g. a trained scikit-learn~\cite{scikit}, Tensorflow~\cite{tensorflow}, or PyTorch model~\cite{pytorch}). 

In the balance of the paper, we present the preliminary design of our solution in section~\ref{sec:approach} and outline avenues for future work in section~\ref{sec:conclusion}. 

\section{Design Goals and Threat Model}
\label{sec:design}
The system should be able to protect the API provider against a cloud vendor(s) trying to steal the model and host an identical service as well as under-reporting of model usage metrics. Similarly, the system should be able to protect honest cloud vendors and API providers from malicious clients who try to cheat by repudiating their inputs and outputs received from the service. Our threat model assumes that cloud vendors are rational and will try to collude with each other to steal the model and host an identical service, as well as under-report model usage statistics.
\section{Solution} 
\label{sec:approach} 

{\mybl We now present the design of a blockchain enabled trustless  API  marketplace that  averts  malicious  behavior of cloud  vendors,  API  providers,  and  API consumers.

\subsection{Solution Design}
At a high level, in our approach, the model encapsulated within the API is split among multiple participating cloud vendors ($CV$) so that no single $CV$ has full knowledge of the deployed model (this is in principle similar to secret sharing~\cite{secret}). 
When the model API is invoked, it is internally executed collaboratively by the respective cloud vendors who successively transform the data to eventually yield the final response to the prediction API. During each step of API call, each participant stakeholder records their transactions on blockchain based on the completed task. }


\subsection{Techniques to split API execution across multiple cloud vendors} 
\label{sec:split} 
Depending on the type of code, software functions, or models encapsulated within an API, its execution steps can easily be split into multiple subcomponents across cloud vendors using several techniques. We present a few techniques that may be used with descriptive and predictive models. 

\subsubsection{Descriptive Models/Software functions} The code related to general software functions or prescriptive models can easily be partitioned into subcomponents so that each subcomponent accepts processed data as input from a previous component, makes new transformations to the data, and outputs it to the next component for further processing. For instance, consider an API commonly used by banks, which computes the credit risk score of a customer based on historical transaction data. If the risk score is computed using $k$ metrics which are combined together within a mathematical formula, then computation of each metric can occur at an individual cloud vendor. The computation of the overall formula can also be hierarchically partitioned among a few cloud vendors so that none of them have knowledge of the mathematical formula, which may be confidential. In this manner each cloud vendor is exposed to only a small portion of the overall computation and does not gain the know how to compute the credit risk score. 

\subsubsection{Predictive Machine Learning Models} 
When an API encapsulates machine learning models, the split of computation for model inference can occur at multiple cloud vendors as follows. For neural network models, each layer or a stack of layers may reside on a cloud vendor, so that the forward propagation step needed for model inference may be executed successively by each cloud vendor and no individual vendor has full knowledge of the neural network model. For decision tree models, each node or a subtree corresponding to the model can reside at a separate cloud vendor, effectively connecting invocations of cloud vendors within a tree structure. Most AI models that yield high accuracy are essentially ensemble models such as random forest, gradient boosted trees, or an ensemble of different models such as SVMs and Neural networks. In such scenarios, each individual model is weak and the ensemble essentially combines several weak models to obtain a stronger model. In this setting, each weak model or a group of weak models may reside on a  cloud vendor. In the proposed alternate approach wherein an  orchestrator invokes a different subset of cloud vendors on each API call, a set of weaker models may be placed at each cloud vendor. The compute step of combining results of multiple models obtained from different vendors (e.g. via any mathematical function such as averaging) can also be split across multiple cloud vendors. 

\subsection{Transactions recorded on blockchain}

{\mybl \noindent\emph{Handshake/Model Distribution and Deployment.~}} Off-chain, an API provider decomposes the model $M$ encapsulated within the API into $n$ subcomponents and containerizes these, effectively yielding $n$ sub-APIs $S_1$,..$S_n$. The provider then enters into individual offline agreements with $n$ cloud vendors and deploys each sub-component $S_{i}$ onto a separate cloud vendor. After decomposition, the provider computes a unique API model signature, for instance the Merkle root~\cite{merkle} of the subcomponents, such that \emph{signature}$(M)=$ \emph{signature}$(S_1..S_{n})$. This model signature is then recorded on the blockchain by the provider. For each participant $p$, let $v(p)$ denote its virtual ID on the blockchain. Thus the API provider knows both real off-chain identities and the virtual identities $v(CV_i)$ $\forall i$ that hold its model. However no two $CV$s that both hold parts of a particular model $M$ know each other's real off-chain identities. In addition to the model signature, the provider records on the blockchain the API id $id(M)$ and the sequence $v(CV_1), ..., v(CV_n)$ in which the subAPIs must be executed by each cloud vendor for an API invocation.  

{\mybl \noindent\emph{API Request Execution.~}} When an API consumer invokes the API $id(M)$ with a user input $I$, the consumer registers $(id(M), I)$ on the blockchain, which notifies all the $CVs$ that hold parts of the model $M$. Thus the cloud vendor $v(CV_1)$ picks up the input $I_1 = I$ from blockchain, executes $S_1$ and obtains the output $O_1$. It then records $(D_1, O_1, hash(S_1))$ on the blockchain. The next cloud vendor in the sequence declared by the provider, $v(CV_2)$ then picks up the input $I_2=O_1$ from the blockchain and repeats the process. The API consumer waits for the cloud vendor $v(CV_n)$ to place its output $O_n$ on the blockchain and eventually picks it up as the final output $O = O_n$ of the API $id(M)$. Figure~\ref{fig:system}(left) presents a schematic of the approach.  

{\mybl \noindent\emph{Dispute Resolution.~}} Smart Contracts are used to ensure the completeness and correctness of execution of sub-components. They can be used to verify the sequence of operations through hash-matches and ensure all $S_1..S_n$ are executed and also verify that each cloud vendor runs the right code for its component through hash-match with $S_i$ provided by model owner.  

%
%
%
%
%
%

\subsection{Alternate approach with orchestrator} 
The previous approach has two drawbacks: (a) the order of execution by cloud vendors remains same across multiple invocations of an API, which can pose risks related to collusion and (b) the blockchain holds input and output data from intermediate sub-APIs which introduces risks related to reverse engineering of sub-components over time. 

We now propose an alternate approach (Figure~\ref{fig:system}(right)), in which $K$ sub-components of a model are stored with overlap across cloud vendors, so that only a subset of $V \subseteq N$ $CV$s are required to invoke a model. In this case, the order and subset of cloud vendors may be chosen randomly for each API call. In addition to the $CV$s, this approach requires the use of an orchestrator who choses a random set of CVs to invoke for each API call. Additionally, the orchestrator passes the input and output data back and forth between the CVs, so that each $CV_i$ (and the orchestrator) records the hash of the input and output data rather than the actual data itself on blockchain i.e. $(hash(D_i), hash(O_i), hash(S_i))$. This approach requires that the orchestrator know the actual off-chain identities (e.g. IP or http address) of the CVs in order to invoke the sub-APIs. The API consumer essentially writes a hash of the input $I$ and the final output $O$ on the blockchain and passes the actual input to the orchestrator and also receives the actual final API output from it. 

\subsection{Collaboration between multiple API providers. }
One of the advantages of the proposed system is that it allows multiple API providers to collaborate together in a trustless setting and offer a joint API without revealing the underlying models to one another. In the previous sections, we considered the setting of an individual API provider who splits the model execution across multiple cloud vendors. However, in practice, different enterprises may want to collaborate in a trustless setting and each build a small component of a larger API and finally monetize the model. For example, a banking firm and an insurance firm may come together to offer a joint API to classify customers for advertising. In this case, both parties use the proposed system to offer a combined API and share revenue from its use in a trustless setting.

\subsection{Security Analysis} 
{\mybl In comparison to centralized marketplaces, the proposed system is resilient against malicious behavior from all three stakeholders.}
\begin{itemize} 
\item \emph{Malicious behavior by cloud vendors. } Since no single $CV$ has access to the full model, any malicious CV cannot copy the deployed model supplied by the model owner and offer a similar service or invoke the model multiple times without reporting to the model owner. Additionally, inaccurate API metrics cannot be reported to API providers as this information is recorded in an immutable ledger. Moreover since cloud vendors do not know each other's off-chain identities, they cannot easily collude to offer a similar service outside of the marketplace. 

\item \emph{Malicious behavior by API providers. } Since blockchain holds evidence of the initial split of model across CVs and the specific modules invoked by each CV along with input and output data, a malicious API provider cannot backtrack or deny the query results of a certain version of the deployed model or provide wrong model to begin with and repudiate that the output has been generated by his model. Additionally, the API provider cannot blame CVs for running inaccurate version of the model components. 

\item \emph{Malicious behavior by API consumers. } Since the blockchain stores evidence of data supplied to the API and to individual model components held by each CV as well as components invoked by each CV, a malicious buyer cannot demand refund by claiming that a wrong output was provided by the service (using a forged output). 

\end{itemize}

\section{Conclusion and Future work} 
\label{sec:conclusion} 
Subscription-driven software licensing mechanisms, wherein providers host software-as-a-service APIs in a cloud environment and get paid on a pay-per-query basis, is now a common approach to monetizing software. Commonly, these  APIs encapsulate confidential and commercially valuable software functions, business intelligence logic or trained machine learning models, which application developers can integrate with data to build software applications. However centralized API marketplaces face challenges related to fairness and trust between all three stakeholders -- cloud Vendors, API providers, and API consumers. Cloud vendors may copy the hosted API models and offer a similar service, thereby diminishing the revenue of API providers. The API providers may backtrack or deny the query results of the deployed API while consumers may demand refund by claiming that a wrong output was provided by the API by potentially forging the output. 

In this work, we presented the design of a blockchain based decentralized API marketplace that allows all stakeholders to work together in a trustless setting without the need for a trusted central entity. The  key  novelty  of  our  work  is  that  instead  of  having  one central cloud vendor store and execute the confidential models encapsulated  within  APIs,  our  system  splits  the execution  of  models  across  multiple cloud vendors  such  that  no  single  vendor  gains  full  knowledge  of the models  invoked  within the  APIs. All stakeholder records their actions on the distributed ledger, so that the system averts malicious behavior and provides evidence of dispute resolution by a trusted arbitrator. 

Future work will implement our system on different blockchain networks (e.g. Ethereum and hyperledger fabric) with the help of chaincode functions and smart contracts as well as study risks related to collusion among participants. An important direction for further investigation is to experimentally compare the performance of centralized API marketplaces with decentralized ones that incur additional costs of recording transactions on the distributed ledger. Another direction is to perform a comparative study of the performance of the proposed system as well as its advantages and disadvantages with techniques that encrypt the hosted API models in order to maintain their confidentiality~\cite{gazelle}. Lastly, secure execution of code and its formal verifiability (as required by each cloud vendor in our solution) is an important area of further study~\cite{iron}. 

\bibliographystyle{IEEEtran}
\bibliography{ref}

\begin{thebibliography}{10}
\providecommand{\url}[1]{#1}
\csname url@samestyle\endcsname
\providecommand{\newblock}{\relax}
\providecommand{\bibinfo}[2]{#2}
\providecommand{\BIBentrySTDinterwordspacing}{\spaceskip=0pt\relax}
\providecommand{\BIBentryALTinterwordstretchfactor}{4}
\providecommand{\BIBentryALTinterwordspacing}{\spaceskip=\fontdimen2\font plus
\BIBentryALTinterwordstretchfactor\fontdimen3\font minus
  \fontdimen4\font\relax}
\providecommand{\BIBforeignlanguage}[2]{{%
\expandafter\ifx\csname l@#1\endcsname\relax
\typeout{** WARNING: IEEEtran.bst: No hyphenation pattern has been}%
\typeout{** loaded for the language `#1'. Using the pattern for}%
\typeout{** the default language instead.}%
\else
\language=\csname l@#1\endcsname
\fi
#2}}
\providecommand{\BIBdecl}{\relax}
\BIBdecl

\bibitem{cloudai}
\BIBentryALTinterwordspacing
Google. (2018) {CloudAI Website}. [Online]. Available:
  \url{https://cloud.google.com/products/ai/}
\BIBentrySTDinterwordspacing

\bibitem{mashape}
\BIBentryALTinterwordspacing
{Kong Inc}. (2008) {Mashape Website}. [Online]. Available:
  \url{https://market.mashape.com/explore}
\BIBentrySTDinterwordspacing

\bibitem{rapidapi}
\BIBentryALTinterwordspacing
RapidAPI. (2010) {RapidAPI Website}. [Online]. Available:
  \url{https://rapidapi.com/}
\BIBentrySTDinterwordspacing

\bibitem{cloudsight}
\BIBentryALTinterwordspacing
CloudSight. (2012) {CloudSight Website}. [Online]. Available:
  \url{https://cloudsight.ai/}
\BIBentrySTDinterwordspacing

\bibitem{azure}
\BIBentryALTinterwordspacing
{Microsoft}. (2018) {Azure Website}. [Online]. Available:
  \url{https://azuremarketplace.microsoft.com/en-us/marketplace/apps}
\BIBentrySTDinterwordspacing

\bibitem{aws}
\BIBentryALTinterwordspacing
{Amazon}. (2018) {AWS Website}. [Online]. Available:
  \url{https://aws.amazon.com/mp/ai/}
\BIBentrySTDinterwordspacing

\bibitem{clarifai}
\BIBentryALTinterwordspacing
{ClarifAI}. (2013) {ClarifAI Website}. [Online]. Available:
  \url{https://clarifai.com/about}
\BIBentrySTDinterwordspacing

\bibitem{algorithmia}
\BIBentryALTinterwordspacing
Algorithmia. (2013) {Algorithmia Website}. [Online]. Available:
  \url{https://algorithmia.com/algorithms}
\BIBentrySTDinterwordspacing

\bibitem{kub}
\BIBentryALTinterwordspacing
{The Linux Foundation}. (2018) {Kubernetes Documentation}. [Online]. Available:
  \url{https://kubernetes.io/docs/concepts/overview/what-is-kubernetes/}
\BIBentrySTDinterwordspacing

\bibitem{danku}
\BIBentryALTinterwordspacing
A.~B. Kurtulmus and K.~Daniel, ``Trustless machine learning contracts:
  {E}valuating and exchanging machine learning models on the ethereum
  blockchain,'' \emph{CoRR}, vol. abs/1802.10185, 2018. [Online]. Available:
  \url{http://arxiv.org/abs/1802.10185}
\BIBentrySTDinterwordspacing

\bibitem{skychain}
\BIBentryALTinterwordspacing
SkyChain. (2017) {SkyChain: The future of Artificial Intelligence in
  Healthcare}. [Online]. Available:
  \url{https://cryptoslate.com/coins/skychain/}
\BIBentrySTDinterwordspacing

\bibitem{neureal}
\BIBentryALTinterwordspacing
Neureal. (2017) {Open-source, peer-to-peer, AI supercomputing, Live data stream
  prediction powered by blockchain, Infinitely scalable}. [Online]. Available:
  \url{https://docs.google.com/document/d/1kOJx7clG2V4TevhgwndRDievXp-VaAciPzjmqGxI0CtA/view}
\BIBentrySTDinterwordspacing

\bibitem{deepchain}
\BIBentryALTinterwordspacing
J.~Weng, J.~Weng, M.~Li, Y.~Zhang, and W.~Luo, ``Deepchain: Auditable and
  privacy-preserving deep learning with blockchain-based incentive,''
  \emph{{IACR} Cryptology ePrint Archive}, vol. 2018, p. 679, 2018. [Online].
  Available: \url{https://eprint.iacr.org/2018/679}
\BIBentrySTDinterwordspacing

\bibitem{ocean}
\BIBentryALTinterwordspacing
{Ocean Protocol Foundation}, ``{Ocean Protocol: A Decentralized Substrate for
  AI Data and Services},'' BigchainDB GmbH and DEX Pte. Ltd, Tech. Rep., March
  2018. [Online]. Available: \url{https://oceanprotocol.com/}
\BIBentrySTDinterwordspacing

\bibitem{ethereum}
\BIBentryALTinterwordspacing
Ehereum. (2014) Ethereum white paper. [Online]. Available:
  \url{https://github.com/ethereum/wiki/wiki/White-Paper}
\BIBentrySTDinterwordspacing

\bibitem{ethereum1}
\BIBentryALTinterwordspacing
{Gavin Woods}. (2018) Ethereum: A secure decentralised generalised transaction
  ledger. [Online]. Available:
  \url{https://ethereum.github.io/yellowpaper/paper.pdf}
\BIBentrySTDinterwordspacing

\bibitem{hyper}
\BIBentryALTinterwordspacing
{Hyperledger Foundation}. (2016) Hyperledger. [Online]. Available:
  \url{https://hyperledger-fabric.readthedocs.io/en/release-1.2/dev-setup/devenv.html}
\BIBentrySTDinterwordspacing

\bibitem{scikit}
F.~Pedregosa, G.~Varoquaux, A.~Gramfort, V.~Michel, B.~Thirion, O.~Grisel,
  M.~Blondel, P.~Prettenhofer, R.~Weiss, V.~Dubourg, J.~Vanderplas, A.~Passos,
  D.~Cournapeau, M.~Brucher, M.~Perrot, and E.~Duchesnay, ``Scikit-learn:
  Machine learning in {P}ython,'' \emph{Journal of Machine Learning Research},
  vol.~12, pp. 2825--2830, 2011.

\bibitem{tensorflow}
\BIBentryALTinterwordspacing
M.~Abadi, A.~Agarwal, P.~Barham, E.~Brevdo, Z.~Chen, C.~Citro, G.~S. Corrado,
  A.~Davis, J.~Dean, M.~Devin, S.~Ghemawat, I.~Goodfellow, A.~Harp, G.~Irving,
  M.~Isard, Y.~Jia, R.~Jozefowicz, L.~Kaiser, M.~Kudlur, J.~Levenberg,
  D.~Man\'{e}, R.~Monga, S.~Moore, D.~Murray, C.~Olah, M.~Schuster, J.~Shlens,
  B.~Steiner, I.~Sutskever, K.~Talwar, P.~Tucker, V.~Vanhoucke, V.~Vasudevan,
  F.~Vi\'{e}gas, O.~Vinyals, P.~Warden, M.~Wattenberg, M.~Wicke, Y.~Yu, and
  X.~Zheng, ``{TensorFlow}: Large-scale machine learning on heterogeneous
  systems,'' 2015, software available from tensorflow.org. [Online]. Available:
  \url{https://www.tensorflow.org/}
\BIBentrySTDinterwordspacing

\bibitem{pytorch}
A.~Paszke, S.~Gross, S.~Chintala, G.~Chanan, E.~Yang, Z.~DeVito, Z.~Lin,
  A.~Desmaison, L.~Antiga, and A.~Lerer, ``Automatic differentiation in
  {PyTorch},'' in \emph{NIPS-W}, 2017.

\bibitem{secret}
A.~Shamir, ``How to share a secret,'' \emph{Commun. ACM}, vol.~22, no.~11, pp.
  612--613, Nov. 1979.

\bibitem{merkle}
R.~Merkle, ``Protocols for public key cryptosystems,'' in \emph{Proc. 1980
  Symposium on Security and Privacy, IEEE Computer Society}.\hskip 1em plus
  0.5em minus 0.4em\relax IEEE Computer Society, 1980, pp. 122--133.

\bibitem{gazelle}
C.~Juvekar, V.~Vaikuntanathan, and A.~Chandrakasan, ``Gazelle: A low latency
  framework for secure neural network inference,'' in \emph{Proceedings of the
  27th USENIX Conference on Security Symposium}, ser. SEC'18, 2018.

\bibitem{iron}
\BIBentryALTinterwordspacing
C.~Hawblitzel, J.~Howell, J.~R. Lorch, A.~Narayan, B.~Parno, D.~Zhang, and
  B.~Zill, ``Ironclad apps: End-to-end security via automated full-system
  verification,'' in \emph{11th {USENIX} Symposium on Operating Systems Design
  and Implementation ({OSDI} 14)}.\hskip 1em plus 0.5em minus 0.4em\relax
  Broomfield, CO: {USENIX} Association, 2014, pp. 165--181. [Online].
  Available:
  \url{https://www.usenix.org/conference/osdi14/technical-sessions/presentation/hawblitzel}
\BIBentrySTDinterwordspacing

\end{thebibliography}

\end{document}